\documentclass{article}
\usepackage{spconf,amsmath,graphicx}

\usepackage{lineno,hyperref}
\usepackage{dsfont}

\usepackage{epsfig,subfigure}
\usepackage{amsmath}
\usepackage{amssymb}
\usepackage{graphicx}
\usepackage{setspace}
\usepackage{multirow}
\usepackage[table]{xcolor}
\usepackage{amsmath,array,graphicx}
\modulolinenumbers[5]
\usepackage[german]{babel}
\usepackage{scalerel}

\usepackage{float}
\usepackage{stfloats}

\usepackage{textcomp}

\usepackage{graphicx}
\usepackage{booktabs}

\usepackage{array,tabularx}
\usepackage{gensymb}

\usepackage[
		     sort&compress]{natbib}
		     		     
\usepackage[table]{xcolor}
\usepackage{makecell}

\usepackage[symbol]{footmisc}

\usepackage{enumitem}

\usepackage{amsmath}

\DeclareMathOperator*{\argmin}{arg\,min}

\makeatletter

\renewcommand\paragraph{\@startsection{paragraph}{4}{\z@}%
                                     {-3.25ex\@plus -1ex \@minus -.2ex}%
                                     {1.5ex \@plus .2ex}%
                                     {\normalfont\normalsize\bfseries}}
                                     
\newcommand{\norm}[1]{\left\lVert#1\right\rVert}

\usepackage{textgreek}

\usepackage{siunitx}
\definecolor{gray_light}{rgb}{0.92,0.92,0.92}

\let\footnotemark

\begin{document}

\title{Tiled sparse coding in eigenspaces for the COVID-19 diagnosis in chest X-ray images}


\def\@name{ \emph{Juan E. Arco$^{1,*}$\thanks{\textsuperscript{*} Corresponding author: jearco@ugr.es}},  \emph{Andr\'es Ortiz$^{2}$}, \emph{Javier Ram\'irez$^{1}$}, \emph{Juan M. G\'orriz$^{1}$}}


\address{\normalsize $^{1}$ Department of Signal Theory, Networking and Communications, Universidad de Granada\\ 
\normalsize $^{2}$ Department of Communications Engineering, Universidad de Malaga \\
}

\maketitle

\begin{abstract}
The ongoing crisis of the COVID-19 (Coronavirus disease 2019) pandemic has changed the world. According to the World Health Organization (WHO), 4 million people have died due to this disease, whereas there have been more than 180 million confirmed cases of COVID-19. The collapse of the health system in many countries has demonstrated the need of developing tools to automatize the diagnosis of the disease from medical imaging. Previous studies have used deep learning for this purpose. However, the performance of this alternative highly depends on the size of the dataset employed for training the algorithm. In this work, we propose a classification framework based on sparse coding in order to identify the pneumonia patterns associated with different pathologies. Specifically, each chest X-ray (CXR) image is partitioned into different tiles. The most relevant features extracted from PCA are then used to build the dictionary within the sparse coding procedure. Once images are transformed and reconstructed from the elements of the dictionary, classification is performed from the reconstruction errors of individual patches associated with each image. Performance is evaluated in a real scenario where simultaneously differentiation between four different pathologies: control \textit{vs} bacterial pneumonia \textit{vs} viral pneumonia \textit{vs} COVID-19. The accuracy when identifying the presence of pneumonia is 93.85\%, whereas 88.11\% is obtained in the 4-class classification context. The excellent results and the pioneering use of sparse coding in this scenario evidence the applicability of this approach as an aid for clinicians in a real-world environment.
\end{abstract}

\begin{keywords}
Pneumonia; COVID-19; sparse coding; machine learning; dictionary.
\end{keywords}

\section{Introduction}
\label{sec:intro}

The ongoing crisis of the COVID-19 (Coronavirus disease 2019) pandemic is still having a terrible effect in health systems of countries worldwide. The World Health Organization (WHO) has confirmed that 4 million people have died due to this disease, whereas the number of infected people rises to 180 million \citep{who}. Despite the distribution of vaccines is highly reducing the number of contagions, it is extremely crucial the early detection of the disease since many people are contagious in the pre-symptomatic period \citep{who2}. Reverse-transcription polymerase chain reaction (RT-PCR) is considered the gold standard for the detection of COVID-19 \citep{pcr1,pcr2,pcr3}. However, the large time need until results are obtained and its relatively low sensitivity can  delay the diagnosis and the subsequent election of the medical treatment. The use of medical imaging can be an alternative solution for the diagnosis of COVID-19, since patients affected by severe COVID-19 usually develop pneumonia. Non-invasive methods such as chest Computed Tomography (CCT) and chest X-ray (CXR) can play a crucial role in the identification of ground glass opacities (GGO) typically associated with COVID-19. Although CCT leads to images with a higher resolution, the low cost of CXR allows that most hospitals have an X-ray machine, resulting in an excellent tool for the diagnosis of this pathology.

Despite medical imaging is extremely useful, it is challenging to discriminate COVID-19 by using only the information that these images offer by several reasons. Findings and abnormalities associated with COVID-19 can be extremely similar than the ones present in other types of pneumonia. The decision of the radiologist can be highly influenced by his/her expertise and there is an overlapping between pneumonia symptoms and lung structures or abnormalities \citep{chandra2020,maduskar2016}. This leads to a manual and slow diagnostic process, with a high inter and intra-observer variability, that can risk the patients' health in situations where the health system is collapsed. For this reason, the use of automatic methods can play a crucial role by alleviating clinicians when the workload is high and serving as a B-reader when the diagnosis is not straightforward. Previous studies have relied on algorithms based on artificial intelligence for developing systems that allow the automatic detection of pneumonia \citep{alizadehsani2021uncertaintyaware,article_covid,hemdan2020covidxnet}. \cite{alizadehsani2021uncertaintyaware} proposed a semi-supervised method based on Sobel edge detection \citep{sobel} and generative adversarial networks \citep{article_gan} to detect the presence of COVID-19. \cite{Ying2020} presented DeepPneumonia, an automatic tool for the identification of COVID-19 patients based on the identification of GGOs.

 Other studies compared different architectures by employing transfer learning on the ImageNet dataset \citep{ajin2017}. \cite{EZZAT2021106742} also adopted a transfer learning method based on a pre-trained DenseNET121. In order to maximize performance, they optimized hyperparameters by using a gravitational search algorithm (GSA), leading to an accuracy of 98.28\%. \cite{ELKORANY2021166405} proposed a deep learning model called COVIDetection-Net for detecting and classifying several types of pneumonia. \cite{arco2020uncertaintydriven} introduced a Bayesian perspective in deep learning models in order to provide not only a classification output but a measure based on uncertainty. Specifically, an ensemble classifier was used and the contribution of each individual classifier to the global system was derived from their uncertainty. This would allow to quantify the reliability of the prediction, which is especially interesting in medical contexts. \cite{arco2021probabilistic} provided a probabilistic alternative based on eigenlungs derived form Kernel Principal Component Analysis (PCA) and ensemble classification. This work focused on CT images instead of CXR ones, and yielded a classification accuracy of 97.86\%. Table \ref{table:summary} provides an overview of recent work focused on the automatic detection of pneumonia.

\begin{table*}[htbp]
\caption{Summary of previous works focused on the automatic identification of pneumonia.}
\label{table:summary}
\centering
\resizebox{\textwidth}{!}{\begin{tabular}{cccccc|}
 \hline
 \rowcolor{gray_light}
  Research work & Dataset & Method & Classification context & Results (\%)\\ 
 \hline
 \cite{alizadehsani2021uncertaintyaware} & 1000 CT scans & GAN model  & Normal \textit{vs} COVID & Acc = 99.95\\
 \cite{arco2020uncertaintydriven} & 6374 CXR images & Bayesian Deep Learning & Normal \textit{vs} Bacterial \textit{vs} Viral  \textit{vs} COVID pneumonia & Acc = 98.06\\
 \cite{arco2021probabilistic} & 513 CT scans & Probabilistic Machine Learning & Normal \textit{vs} COVID& Acc = 97.86 \\
 \cite{cong2020} & 137 CT scans & 3D-Resnet-10 & Severe \textit{vs} Critical COVID & AUC = 90.9 \\
 \cite{article_covid} & 4356 CT scans & COVNet & Normal \textit{vs} COVID & AUC = 96\\
 \cite{Panetta2021AutomatedDO} & 2905 CXR images & Fibonacci patterns & Normal \textit{vs} COVID & Acc = 99.78\\
 \cite{9387536} & 13962 CXR images &  DeepDRR & Normal \textit{vs} COVID & Acc = 94\\ 
 \cite{Narin2021AutomaticDO} & 3141 CXR images &  Resnet-50 & Normal \textit{vs} COVID & Acc = 96.1\\ 
 \cite{9144185} & 3487 CXR images &  DenseNet-201 & Normal \textit{vs} Bacterial \textit{vs} Viral pneumonia& Acc = 97.94\\ 
 \cite{antonios2020} & 1428 CXR images &  VGG19 & Normal \textit{vs} COVID \textit{vs} Bacterial pneumonia& Acc = 98.75\\ 
 \cite{hoon2020} & 3993 CXR images &  Resnet-50 & Normal \textit{vs} COVID \textit{vs} Other pneumonia& Acc = 99.87\\ 
 \cite{tabik2020} & 852 CXR images &  COVIDNet & Normal \textit{vs} COVID & Acc = 97.72\\ 
 \cite{mustafa2021} & 3487 CXR images &  CheXNet & Normal \textit{vs} Bacterial \textit{vs} Viral pneumonia & Acc = 97.8\\ 
 \cite{OZTURK2020103792} & 1142 CXR images &  DarkNet & Normal \textit{vs} COVID \textit{vs} Viral pneumonia & Acc = 87.02\\ 
 \cite{MISHRA2021572} & 400 CT scans &  VGG16 & Normal \textit{vs} COVID & Acc = 99\\ 
 \cite{KASSANIA2021867} & 234 CT scans &  DenseNet-121 & Normal \textit{vs} COVID & Acc = 99\\ 
 \cite{Ibrahim2021AbnormalityDA} & 1110 CT scans &  COV-CAF & Normal \textit{vs} COVID & Acc = 97.76\\  
 \cite{c1dfefe954874649ba99338988a5b826} & 1164 CT scans &  CCSHNet & Normal \textit{vs} COVID \textit{vs} Pneumonia \textit{vs} Tuberculosis & Acc = 96.46\\ 
 \cite{jangam2021} & 4886 CXR images &  Ensemble Deep Learning & Normal \textit{vs} COVID& Acc = 99.8\\ 
 \cite{RAHIMZADEH2021102588} & 63849 CT scans &  ResNet-50V2 & Normal \textit{vs} COVID& Acc = 99.49\\ 
\hline
\end{tabular}}
\end{table*}

The performance of deep learning approaches highly depends on the size of the dataset available. When the number of samples is high, methods based on deep learning usually learn the main features that characterize the different classes to distinguish from. When data is limited, deep networks could not learn the relationship between samples and labels. Thus, the use of methods derived from deep learning alternatives would be restricted to scenarios where thousands of images are available. Despite the rising of public datasets, the requirements needed for the correct use of deep learning approaches can not always be met. In this work, we provide an alternative solution based on sparse coding in order to maximize the identification of the clinical symptoms associated with COVID-19 from CXR images. Each one is divided into a number of squared tiles (also known as patches), and features are extracted by using PCA. The most relevant components are then used to build a dictionary, which is the basis of sparse coding. Images are coded and reconstructed according to the elements of the dictionary, obtaining a reconstruction error for each individual patch. Finally, the reconstruction errors of all patches contained in the images are then entered into a classifier, which decides the presence (or not) of pneumonia. Performance of this classification framework is evaluated in a range of scenarios of incremental difficulty: from a control \textit{vs} pneumonia patients to a multiclass context where differentiating between four pathologies: controls \textit{vs} bacterial pneumonia \textit{vs} viral pneumonia \textit{vs} COVID-19 pneumonia. The main contributions of our work can be summarized as follows:

\begin{itemize}
\item{The pioneering use of sparse coding allows a novel and rapid approach for the diagnosis of pneumonia.}
\item{The application of PCA optimizes the informativeness of the elements included in the dictionary.}
\item{The division into patches allows the spatial identification of the pathology and its cause (bacteria, virus, COVID-19).}
\item{Our approach offers an alternative to deep learning for scenarios where the number of samples is limited.}
\end{itemize}

\section{Material}
\label{sec:materials}

\subsection{Dataset}
\label{subsec:dataset}
\begin{figure*}
\centering
\includegraphics[width=0.8\textwidth]{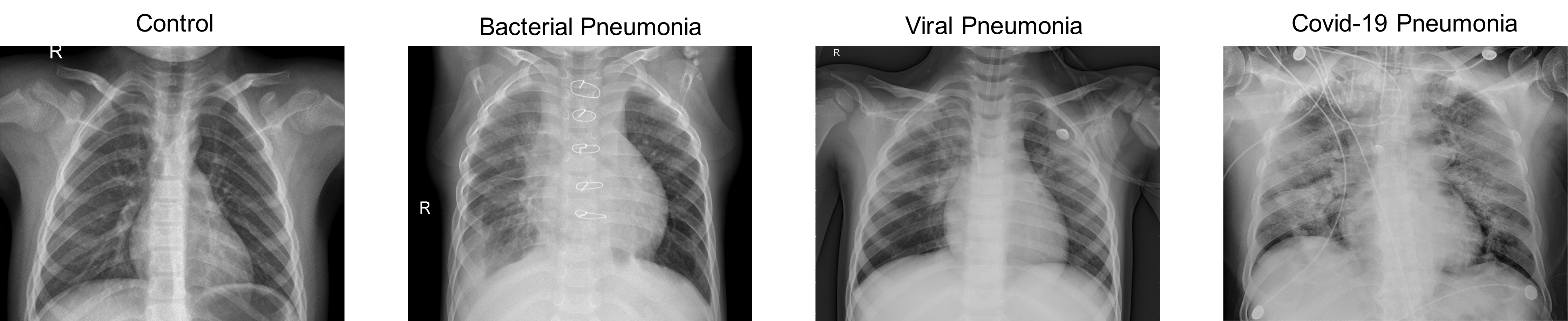}
\caption{From left to right, CXR image of a control, bacterial pneumonia, viral pneumonia and COVID-19 pneumonia. Note some clear artifacts in Covid-19 image.}
\label{fig:figurauno}
\end{figure*}

We have used the dataset available in \cite{kaggle} for controls and patients who suffered from a bacterial or a non-COVID19 pneumonia. According to the information described in \cite{kermany2018}, the CXR images were selected from retrospective cohorts of pediatric patients of one to five years old from Guangzhou Women and Children's Medical Center, Guangzhou. All CXR images were acquired as part of patient's routines clinical care. Institutional Review Board (IRB)/Ethics Committee approvals were obtained. The work was conducted in a manner compliant with the United States Health Insurance Portability and Accountability Act (HIPAA) and was adherent to the tenets of the Declaration of Helsinki. \cite{kermany2018} collected and labeled a total of 5856 CXR images from children, including 4273 characterized as depicting pneumonia and 1583 normal. From those patients diagnosed with pneumonia, 2786 were labeled as bacterial pneumonia, whereas 1487 were labeled as viral pneumonia. The dataset containing COVID-19 patients is available in \cite{kaggle1} and includes 576 CXR images from adults. Figure \ref{fig:figurauno} shows the CXR image from a control (CTL), and a patient suffering from a bacterial (BAC), a viral (VIR) and a COVID19 (CVD19) pneumonia.

\subsection{Image preprocessing}
\label{subsec:prepro}
When working with medical images, it is extremely important to apply a preprocessing in order to improve the subsequent classification performance.This is especially important in CXR images, where low X-ray radiation and movement during image acquisition result in noisy and low-resolution images. This preprocessing also adapts images to the needs of the neural network. In order to mitigate computational and memory limitations, we downsampled the input images to obtain a final map of size 224x224. We also performed an intensity normalization procedure for each individual image based on standardization. Each image was transformed such the resulting distribution has a mean (\begin{math} \mu \end{math}) of 0 and a standard deviation (\begin{math} \sigma \end{math}) of 1, as follows:

\begin{equation}
I' = \frac{I-\mu}{\sigma}
\label{eq:clahe3}
\end{equation}

\noindent where \begin{math} I \end{math} is the original image and I' is the resulting one.

\section{Methods}
\label{sec:methods}

\subsection{Sparse coding}
\label{subsec:sparse}

The idea behind this technique is derived from the way the primary visual cortex in human brain works. The number of neurons in this brain region is much higher than the number of receptor cells in the retina, which suggests that a sparse code is used to efficiently represent natural scenes \citep{cortex1,cortex2,cortex3}. Sparse coding relies on the assumption that data can be represented in terms of a linear combination of basis elements \citep{sparse1}. Consider a number of samples of class \begin{math} i \end{math}, \begin{math} A_{i} = [\mathbf{v}_{i,1}, \mathbf{v}_{i,2}, \ldots, \mathbf{v}_{i,n_{i}}] \in \mathds{R}^{m \times n_i} \end{math}. A new sample \begin{math} y \in \mathds{R}^{m}  \end{math} can be approximated by the linear span of the initial number of samples as follows:

\begin{equation}
\label{eq:eq1}
\mathbf{y} = \alpha_{i,1} \mathbf{v}_{i,1} + \alpha_{i,2} \mathbf{v}_{i,2} + \cdots + \alpha_{i,n_{i}} \mathbf{v}_{i,n_{i}}
\end{equation}

\noindent where \begin{math} \alpha_{i,j} \in \mathds{R}, j=1,2,\ldots, n_{i} \end{math} is the coefficient vector.

\begin{math} \end{math}

\subsection{Creation of the dictionary}
\label{subsec:dictionary}

Sparse coding allows the representation of a signal in terms of a linear combination of a few atoms of a dictionary matrix. The main advantage of this technique is that a complex signal can be represented in a very concise manner. A crucial aspect in this process is the way the dictionary is built. The simplest solution is to use each individual image as a dictionary entry. Thus, the size of the dictionary would be exactly the same than the number of images available. This can be problematic when the size is too high for two main reasons. First, a dictionary with a high number of atoms would lead to a slow process when transforming and reconstructing the images because of the matrices multiplication this process relies on. Second, employing features in the original space can be suboptimal when trying to maximize the differences in the representation of images of different classes. In this work, we propose a patches-based method relying on Principal Component Analysis (PCA) for the creation of the dictionary. Briefly, original images are divided into patches of a fixed size, storing then each individual patch by columns in a matrix. PCA \cite{pca1,pca2,pca3} is then applied to this matrix. Given a set of \begin{math} N \end{math} samples \begin{math} \mathbf{x}_k \end{math}, \begin{math} \mathbf{x}_k  = [\mathbf{x}_{k1}, \ldots , \mathbf{x}_{kn} ] \in  \mathbb{R}^n\end{math}, the aim of PCA is to find the projection directions that maximize the variance of a subspace \cite{lopez}. Thus,  vector \begin{math} \mathbf{x} \end{math} is projected from the input space, \begin{math}\mathbb{R}^n \end{math}, to a high-dimensional space,  \begin{math} \mathbb{R}^f \end{math}. In the new feature space, \begin{math} \mathbb{R}^f \end{math}, the eigenvalue problem can be described as follows:

\begin{equation}
\label{eq:pca1}
C^{\Phi} \mathbf{w}^{\Phi} = \lambda \mathbf{w}^{\Phi}
\end{equation}

\noindent where \begin{math} C^{\Phi} \end{math} is a covariance matrix. All the solutions \begin{math} \mathbf{w}^\Phi \end{math} with \begin{math} \lambda \neq 0\end{math} are in the transformed space \begin{math} \Phi(\mathbf{x}_1,\ldots, \Phi(\mathbf{x}_N)) \end{math}, and there exist coefficients 
\begin{math} \alpha_i \end{math} such that:

\begin{equation}
\label{eq:pca2}
\mathbf{w}^\Phi = \sum_{i=1}^{N} \alpha_{i} \Phi(\mathbf{x}_i)
\end{equation}

Defining an \begin{math} N x N\end{math} matrix \begin{math} K \end{math} by 
\begin{equation}
\label{eq:pca3}
K_{ij} = k(\mathbf{x}_i,\mathbf{x}_j) = \Phi(\mathbf{x}_i) \cdot \Phi(\mathbf{x}_j)
\end{equation}

\noindent the PCA problem becomes:

\begin{equation}
\label{eq:pca4}
N \lambda K \boldsymbol{\alpha} = K^{2} \boldsymbol{\alpha} \equiv N\lambda \boldsymbol{\alpha} = K \boldsymbol{\alpha}
\end{equation}

\noindent where \begin{math} \boldsymbol {\alpha} \end{math} denotes a column vector with entries \begin{math} \alpha_{1} \ldots \alpha_{N}\end{math} \cite{kernel_pca2}.

Finally, vectors in the high-dimensional feature space are projected into a lower dimensional spanned by the eigenvectors \begin{math} \mathbf{w}^\Phi \end{math}. Given a sample \begin{math} \mathbf{x} \end{math} whose projection is \begin{math} \Phi(\mathbf{x}) \end{math} in \begin{math} \mathbb{R}^f \end{math}, the projection of  \begin{math} \Phi(\mathbf{x})\end{math} onto the eigenvectors \begin{math} \mathbf{w}^\Phi \end{math} is the nonlinear principal components corresponding to \begin{math} \Phi \end{math}, as follows:

\begin{equation}
\label{eq:pca6}
\mathbf{w}^{\phi} \cdot  \Phi(\mathbf{x}) = \sum_{i=1} ^{N} \alpha_{i}(\Phi(\mathbf{x}_{i}) \Phi(\mathbf{x})) = \sum_{i=1}^{N} \alpha_{i}K(\mathbf{x}_i,\mathbf{x})
\end{equation}

We divided the images into patches instead of using complete images because of the patterns associated with pneumonia are usually located in small regions. The sparse coding stage can be summarized as follows:

\begin{itemize}
\item{Division of the images into patches. We did this operation because patterns associated with pneumonia are usually located in small regions, which means that it could be better to analyze individual to perform the diagnosis.}
\item{Creation of a preliminary dictionary where each column corresponds to individual patches. The size of the resulting matrix will be \begin{math} M x N\end{math}, where \begin{math} M \end{math} is the number of voxels contained in each patch and \begin{math} N \end{math} is the product of the number of patches contained in each image and the total number of images.}
\item{Application of PCA to this matrix to obtain an optimum dictionary by maximizing the variance of the projected components. The eigenvectors will be the elements of the dictionary, so that their size will be derived from the number of principal components preserved.}
\item{Computation of the decomposition coefficient vector \begin{math} \hat{\alpha} \end{math} by solving the L1-norm minimization problem by sparse coding:

\begin{math} \hat{\alpha} = \argmin_{\alpha} \norm{\alpha}_{1} \end{math}  \text{subject to} \begin{math} \norm{\mathbf{X} \alpha-\mathbf{y}}_{2} \leq \epsilon \end{math} where \begin{math} \mathbf{X} \end{math} is the dictionary matrix and \begin{math} \mathbf{y} \end{math} is the test sample to be transformed.}

\item{Estimation of the sparse reconstruction error for each patch \begin{math} \mathbf{y} \end{math} by employing the sparse coefficients associated \begin{math} \hat{\alpha} \end{math}}:
\begin{math} e = \norm{\mathbf{X} \hat{\alpha}} -\mathbf{y}\end{math}
\end{itemize}

\begin{figure*}
\centering
\includegraphics[width=0.9\textwidth]{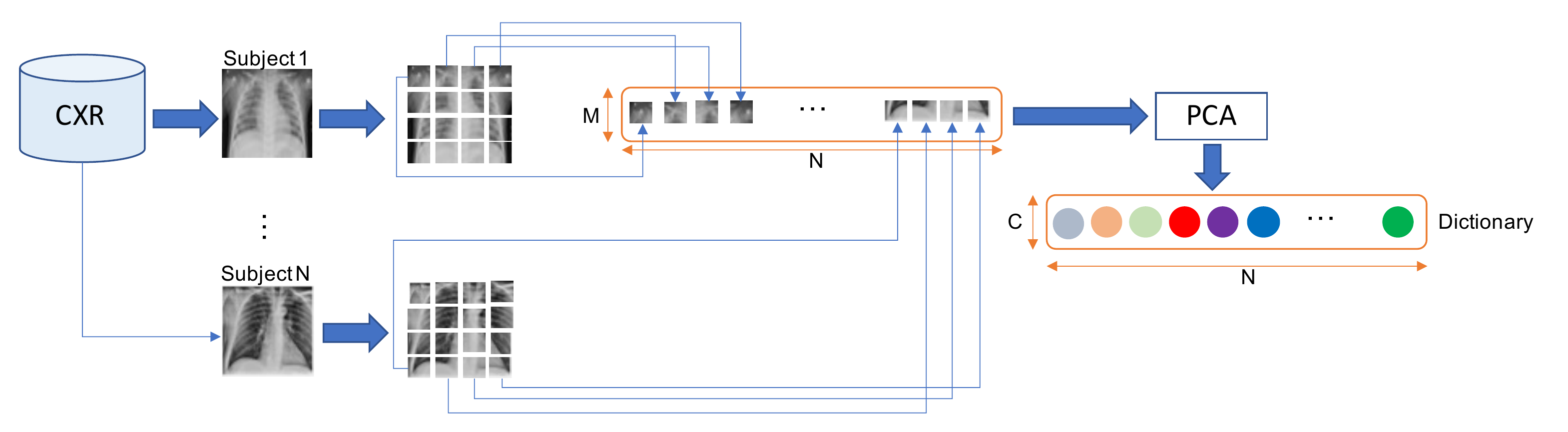}
\caption{Process for the creation of the dictionary. First, images are divided into patches of equal size. These patches are organized by columns into a matrix. A PCA is then used to maximize the variance of the projected components. The resulting ones form the final dictionary.}
\label{fig:dictionary}
\end{figure*}

These operations can be seen as a feature extraction prior to the classification. Once the reconstruction errors are computed for all images, a new dataset is generated with as many features as patches and as many samples as images. Figure \ref{fig:dictionary} depicts a schematic representation of the dictionary generation. The aim at this point is that the classifier finds the relationship between features and samples. We employed an SVM algorithm for binary classification (CTL \textit{vs} PNEU) and a Random Forest (RF) classifier for multiclass classification.

\subsection{Classification}
\label{subsec:classify}

\subsubsection{Support Vector Machines}
\label{subsubsec:svm}

The resulting reconstruction errors were then entered as input of the classification process, which was based on an SVM classifier with a linear kernel. This approach employs the hyperplane with the maximum separation between classes to distinguish between them. This separation is known as margin, and the nearest data points are usually termed support vectors. From a mathematical perspective, it is possible to specify a linear SVM classification rule \begin{math} f \end{math} by a pair of \begin{math} (\mathbf{x}, \mathbf{x})\end{math}, as follows:

\begin{equation}
\label{eq:svm1}
f(\mathbf{x}_i) = \langle \mathbf{w}, \mathbf{x}_i \rangle + b
\end{equation}

\noindent where \begin{math} \mathbf{w} \end{math} is the weight vector, \begin{math} \mathbf{x}_i \end{math} is the feature vector and \begin{math} b \end{math} is the error term. Thus, a point \begin{math} x \end{math} is classified as positive if \begin{math} f(x) >0 \end{math} or negative if \begin{math} f(x)<0 \end{math}. The maximum distance between the two classes is obtained by solving the optimisation problem described in \cite{svm_boser}:

\begin{equation}
\label{eq:svm2}
\begin{gathered}
\frac{1}{2} \lVert{\mathbf{w}\rVert}^2 + C \sum_{i}\xi_{i}  \hspace{1 cm} \text{subject to}\\
y_{i} (\langle \mathbf{w},\mathbf{x}_i \rangle + b) \geq 1-\xi_i \hspace{1 cm} \forall_i\xi_i \geq 0 \hspace{0.2 cm} \forall_i
\end{gathered}
\end{equation}

\noindent where \begin{math} C \end{math} is usually known as penalty for misclassification, or cost parameter. The solution to the optimisation problem can be written as:

\begin{equation}
\label{eq:svm3}
\mathbf{w} = \sum_{i=1}^{n} y_{i}\alpha_{i}\mathbf{x}_i
\end{equation}

\noindent after applying the Lagrangian multipliers. Substituting the value of \begin{math} \mathbf{w} \end{math} in Equation \ref{eq:svm1}, it is possible to rewrite the decision function in its dual form as:

\begin{equation}
\label{eq:svm4}
f(\mathbf{x}_i) = \sum_{i=1}^{n}\alpha_{i}K(\mathbf{x},\mathbf{x}_i)+b
\end{equation}

\noindent where \begin{math} \alpha_i \end{math} and \begin{math} b \end{math} represent the coefficients to be learned from the examples and \begin{math} K(\mathbf{x},\mathbf{x}_i) \end{math} is the kernel function that characterizes the similarity between samples \begin{math}\mathbf{x}\end{math} and \begin{math}\mathbf{x}_i \end{math}.

Since classes were unbalanced (e.g. the number of pneumonia patients was higher than controls), we incorporated the weights of the classes into the cost function in order to the majority class does not contribute more than the minority one.

\subsubsection{Random Forest}
\label{subsubsec:rf}
For the multiclass classification (controls \textit{vs} different types of pneumonia), the reconstruction errors of the different patches of the images were entered into a Random Forest classifier. RF is an ensemble method that combines a number of decision trees in order to improve the performance of individual classifiers. The trees are built from \textit{k} random vectors, \begin{math}\mathbf{\Theta}_{k} \end{math}, which are independent of the past random vectors \begin{math}\mathbf{\Theta}_{1},\mathbf{\Theta}_{2},\mathbf{\Theta}_{3},\dots,\mathbf{\Theta}_{k-1} \end{math} but with the same distribution. The process developed by \cite{breiman} employs bagging for generating each random vector \begin{math} \mathbf{\Theta} \end{math} as the \begin{math} N \end{math} observations randomly drawn from the training set. Once a large number of trees is generated \begin{math} \{h(\mathbf{x},\Theta_{k}),k=1,\dots \}\end{math}, each one of them casts a vote for the most popular class at input \begin{math} \mathbf{X} \end{math}. The final decision of the classifier is determined by a majority vote of the trees.

One important feature of RF classifiers is related to its convergence and generalization error \citep{Yang2008,breiman}. Given a set of classifiers \begin{math} h_{1}(\mathbf{x}), h_{2}(\mathbf{x}), \dots, h_{k}(\mathbf{x})\end{math}, and a training set randomly drawn from the distribution of a random vector (X,Y), the margin function is defined as:

\begin{equation}
\label{eq:rf}
mg(\mathbf{X},Y) = av_{k} I(h_{k}(\mathbf{X})=Y)-max_{j\neq Y} av_{k}I(h_{k}(\mathbf{X})=j)
\end{equation}

\noindent where \begin{math} \mathbf{X} \end{math} is the input metric, \begin{math}av_{k} \end{math} is the average number of votes at \begin{math} \mathbf{X} \end{math}, \begin{math} Y \end{math} for the corresponding class and \begin{math} I(\cdot) \end{math} is the indicator function. The margin is a measure about the extent to which the average number of votes at \begin{math} \mathbf{X},Y\end{math} for the right class exceeds the average vote for any other class. Thus, the larger the margin, the more confidence in the classification. The generalization error es given by:

\begin{equation}
\label{eq:rf2}
PE^{*} = P_{\mathbf{X},Y} (mg(\mathbf{X},Y) <0)
\end{equation}

\noindent where \begin{math} P_{\mathbf{X},Y}\end{math} indicates that the probability is over the \begin{math} \mathbf{X},Y \end{math} space. According to Theorem 1.2 in \cite{breiman}, as the number of trees increases, for almost surely all sequences \begin{math} \Theta_{1},\dots PE^{*}\end{math} converge to:

\begin{equation}
\label{eq:rf3}
P_{\mathbf{X},Y}(P_{\Theta}(h(\mathbf{X},\Theta)=Y)-max_{j\neq Y} P_{\Theta}(h(\mathbf{X},\Theta)=j)<0)
\end{equation}

Equation \ref{eq:rf3} shows that for a large number of trees, it follows the Strong Law of Large Numbers. This explains that random forests do not overfit as more trees as added to the ensemble since they produce a limiting value of the generalization error. A crucial aspect of RF models is related to the two randomized procedures applied for building the trees \citep{Yang2008}. For a number of cases in the training set, \begin{math} N,\end{math} and a number of variables in the classifier,\begin{math} M,\end{math} the number of input variables used for determining the decision at a node of the tree will be given by \begin{math} m \end{math}. This number should be much less than \begin{math} M \end{math} \begin{math} (m <<M) \end{math}.  Briefly, from the training set, \begin{math} N \end{math} samples are randomly selected with replacement as the new training set. For each node of the tree, \begin{math} m \end{math} of the \begin{math} M \end{math} variables on which to base the decision at that node are randomly selected. The best split based on these \begin{math} m \end{math} variables is computed, and each tree is fully grown and not pruned.

\begin{figure*}
\centering
\includegraphics[width=0.9\textwidth]{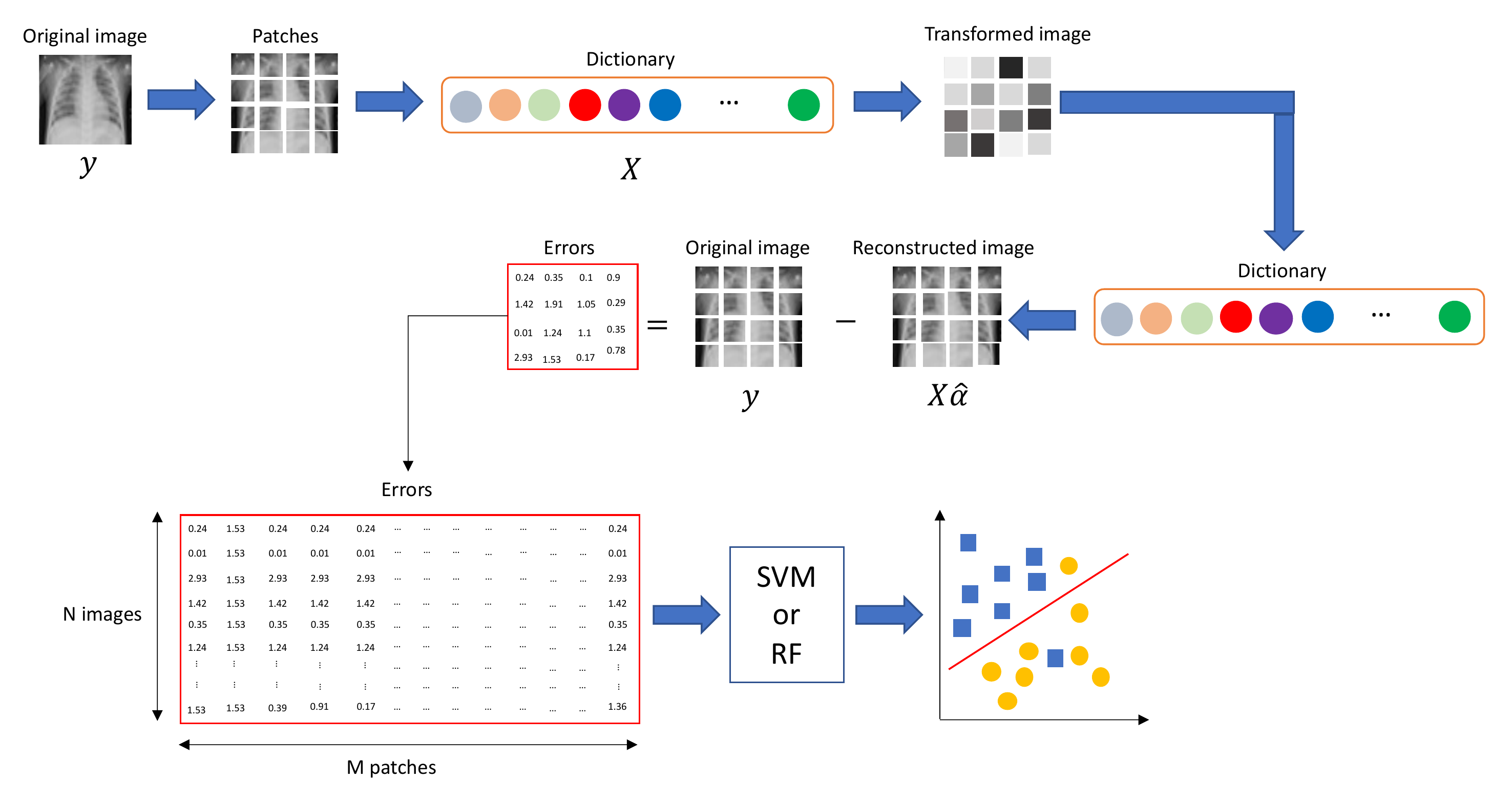}
\caption{Scheme of the classification pipeline. After building the dictionary, the images are transformed by employing sparse coding. The transformed ones are then projected back by reconstruction. The errors of this process are computed as the difference between the original and the transformed images. The resulting errors obtained in each individual patch are then used as the input of the classifier, leading to the label prediction.}
\label{fig:classification}
\end{figure*}

Figure \ref{fig:classification} shows a schematic representation of the entire classification process. Once the dictionary is built, images are transformed and reconstructed, leading to an error value when compared the transformed with the original images. The reconstruction errors for each individual patch are then entered into the classifier, which assigns the final label prediction.

\subsection{Performance evaluation}
\label{subsec:performance}

For all experiments, a 5-fold stratified cross-validation scheme was used to estimate the generalization ability of our method \citep{Kohavi95}. We evaluated the performance of the classification frameworks in terms of the following parameters from the confusion matrix:

\begin{align}
\label{eq:metrics}
\begin{split}
Bal \hspace{0.1cm} Acc = \frac{1}{2} \Big( \frac{TP}{P} + \frac{TN}{N}\Big) \hspace{0.3cm} Sens = \frac {T_{P}}{T_{P}+F_{N}}  
\\
Spec = \frac {T_{N}}{T_{N}+F_{P}} \hspace{0.5cm} AUC = \frac{1}{2} \Big( \frac{TP}{P} + \frac{TN}{N}\Big) 
\\
Prec = \frac {T_{P}}{T_{P}+F_{P}}  \hspace{0.5cm} F1-score = \frac{2 \times Prec \times Sens}{Prec + Sens}
\end{split}
\end{align}

\noindent where \begin{math}  T_{P} \end{math} is the number of pneumonia patients correctly classified (true positives), \begin{math} T_{N} \end{math} is the number of control patients correctly classified (true negatives), \begin{math} F_{P} \end{math} is the number of control subjects classified as pneumonia (false positives) and \begin{math} F_{N} \end{math} is the number of pneumonia patients classified as controls (false negatives). We also employed the area under the curve ROC (AUC) as an additional measure of the classification performance \citep{auc1,auc2}. In the multiclass scenario, the information derived from parameters such as Sens or Spec can not be easily interpreted. In this context, we employed a method based on a multi-class One-vs-One scheme to compare every unique pairwise combination of classes \citep{multi2001}. The multiclass-AUC was computed by averaging the results obtained for each individual comparison. Moreover, a multiclass version of the balanced accuracy was computed, as follows:

\begin{equation}
\label{eq:multi_metrics}
Multiclass \hspace{0.1cm} Bal \hspace{0.1cm} Acc = \frac{1}{M} \sum_{m=1}^{M} \frac{r_{m}}{n_{m}}
\end{equation}

\noindent where \begin{math} M \end{math} is the number of classes, \begin{math}n_m \end{math} is the number of samples belonging to class \begin{math} m \end{math} and \begin{math} r_m \end{math} is the number of samples belonging to class \begin{math} m \end{math} that are accurately predicted.

\section{Evaluation}
\label{sec:eval}

\subsection{Experimental setup} 
\label{subsec:setup}
In this work we propose a classification framework to identify the patterns associated with pneumonia from CXR images. To do so, we define two experiments:

\begin{itemize}[leftmargin=*]
\item{\textbf{Experiment 1: Binary Classification} to distinguish between different groups in three contexts: \textbf{CTL \textit{vs} PNEU}}, which includes all images labelled as CTL and PNEU regardless of the type of pneumonia; \textbf{BAC \textit{vs} VIR}, which divides the images from patients diagnosed from pneumonia according to the cause of the disease (bacterial or viral); \textbf{VIR \textit{vs} CVD19} for viral pneumonia. In the last context, the aim was to identify whether viral pneumonia was caused by COVID-19 or not. In this first experiment, the resulting features from the sparse coding phase were then entered into a linear SVM classifier. We varied the  size of the patches and the number of resulting components from PCA during the construction of the dictionary in order to evaluate its influence in the performance. The parameters associated with this algorithm were optimized in a grid-search process within the training phase.

\item{\textbf{Experiment 2: Multiclass Classification} by using an RF classifier in order to distinguish between the four different pathologies contained in the database. This algorithm combines the decisions of individual trees to obtain the final diagnosis of the patient. The process for building the dictionary in addition to the classification framework are identical to Experiment 1 except the aforementioned change in the classier.}
\end{itemize}

\section{Results}
\label{sec:results}

\section{Results}
\label{sec:results}

\begin{table*}[htbp]
\caption{Balanced accuracies (and their deviations) of the classification approach proposed in this work in the different contexts evaluated. Patch size is given in pixels.}
\label{table:results1}
\centering
\resizebox{\textwidth}{!}{\begin{tabular}{cccccccccc}
 \hline
 \rowcolor{gray_light}
 \multicolumn{10}{c}{Controls \textit{vs} Pneumonia}\\
 \hline
  \multicolumn{10}{c}{Number of PCA components}\\ 
 \hline
 Patch size & N=1& N=2& N=3 & N=4 & N=5 & N=6 & N=7 & N=8 & N=9\\
 \hline
7x7  & 79.45\textpm 1.78 & 82.38\textpm 1.49 &  87.77\textpm 1.46 & 88.09\textpm 1.23 & 87.25\textpm  1.29&  88.16\textpm  1.67 & 88.11\textpm  1.86 & 87.56\textpm 1.91& 87.35\textpm 1.92\\
8x8 & 80.54\textpm 1.98 &83.25\textpm 1.57 &88.49\textpm  1.87 & 88.97\textpm  1.72 & 86.76\textpm 1.98 & 87.02\textpm 1.61 & 88.29\textpm 1.52 & 88.08\textpm 1.73 & 88.14\textpm 1.87\\
14x14 & 81.29\textpm 1.72 & 89.31\textpm 1.08 & 92.13\textpm 1.26  & 92.14\textpm 1.13 & \textbf{93.85\textpm 1.27} & 92.11\textpm 1.13 & 92.17\textpm 0.87 & 91.99\textpm 1.09 & 92.06 \textpm 1.06\\
16x16 & 82.35\textpm 1.23 & 87.99\textpm 1.48 & 91.75\textpm 1.00 & 91.76 \textpm 0.91 & 91.55\textpm 1.04 & 91.65 \textpm 1.13 & 91.50\textpm 1.32 & 91.87\textpm 1.35 & 91.76\textpm 1.21\\
28x28 & 79.94\textpm 1.54 & 85.19\textpm 0.78 & 87.56\textpm 2.38 & 88.05\textpm 1.21 & 88.37\textpm 1.27 & 89.10\textpm 1.42  & 89.20\textpm 0.93 & 89.18 \textpm 1.25 & 89.32 \textpm 0.99\\
32x32 & 78.45\textpm 1.71 & 84.10\textpm 1.91& 88.56\textpm 2.84 & 88.92\textpm 1.71 & 89.75\textpm 2.05 & 89.44\textpm 1.38 & 84.73\textpm 4.98 & 86.61\textpm 4.54 & 88.42\textpm 1.71\\
56x56 & 65.39\textpm 2.15 & 69.47\textpm 1.77 & 72.85\textpm 6.78 & 72.49\textpm 11.91 & 74.95\textpm 3.97 & 73.63\textpm 4.49 & 74.07\textpm 8.73  & 72.73\textpm 7.02 & 75.78\textpm 8.32\\
 \hline
 \rowcolor{gray_light}
 \multicolumn{10}{c}{Bacterial \textit{vs} Viral pneumonia}\\
 \hline
 \multicolumn{10}{c}{Number of PCA components}\\ 
 \hline
 Patch size & N=1 & N=2 & N=3 & N=4 & N=5 & N=6 & N=7 & N=8 & N=9\\
 \hline
7x7  & 75.43\textpm 2.15&80.28\textpm 1.55 & 84.55\textpm 1.11 & 84.19\textpm 1.44 & 83.07\textpm 1.36 & 83.78\textpm 1.41 & 84.12 \textpm 1.55 &83.35 \textpm 1.67& 84.36\textpm 1.03 \\
8x8 & 75.91\textpm 1.90&81.29\textpm 1.14 & 85.69\textpm  1.04 & 85.35\textpm  1.12 & 86.83\textpm 0.93 & 84.98\textpm 1.58 & 85.09 \textpm 1.09 & 85.61\textpm 1.02 & 85.77 \textpm 0.89\\
14x14 & 77.12 \textpm 1.54& 84.51\textpm 1.59  & 88.56 \textpm  0.95 & 88.75 \textpm 0.76 & 88.48\textpm  0.59 & 88.10\textpm 0.67 & 87.97\textpm  0.52 & 88.79\textpm 0.67& 88.32\textpm 0.87\\
16x16 &76.59\textpm 1.62 &83.35\textpm 1.33 &  88.34\textpm 0.86 & 88.56\textpm 0.34 & 88.69\textpm 0.55 & 88.53\textpm 0.58 & 88.49\textpm 0.79 & \textbf{ 88.85\textpm 0.91} & 88.60\textpm 0.86\\
28x28 & 71.36\textpm 2.53& 77.97\textpm 1.47 & 80.72\textpm 0.82 & 80.48\textpm 0.69 & 80.35\textpm 1.05 & 80.51\textpm 0.73 & 80.77\textpm 0.72 & 80.49\textpm 0.66 & 80.73\textpm 1.15\\
32x32 & 69.92 \textpm 2.48 & 76.29\textpm 1.58 & 78.42\textpm 1.09 & 78.71\textpm 1.13 & 79.03\textpm 0.87 & 78.77\textpm 1.65 & 78.19\textpm 1.21 & 78.80\textpm 1.06 & 78.48\textpm 1.53\\
56x56 & 62.36\textpm 2.24& 68.12\textpm 1.21& 70.37\textpm 1.40 & 71.06\textpm 1.29 & 70.68\textpm 1.12 & 70.45\textpm 1.27 & 71.38\textpm 0.39 & 70.71\textpm 1.55 & 70.30\textpm 1.27\\
 \hline
 \rowcolor{gray_light}
 \multicolumn{10}{c}{Viral \textit{vs} COVID19 pneumonia}\\
 \hline
 \multicolumn{10}{c}{Number of PCA components}\\ 
 \hline
 Patch size & N=1& N=2 & N=3 & N=4 & N=5 & N=6 & N=7 & N=8 & N=9\\
 \hline
7x7  & 77.28\textpm 1.75 & 81.23\textpm 1.88 & 86.78\textpm 1.93 & 86.98\textpm  1.83& 87.29\textpm 2.14 & 87.37\textpm 2.07  & 87.81\textpm 1.67 & 86.54\textpm 1.99 & 86.23 \textpm 1.86\\
8x8 & 78.03\textpm 1.72 &86.65\textpm 1.43 & 89.94\textpm 1.58  & 90.03\textpm 1.42 & 88.76\textpm 1.27 & 89.47\textpm 1.45 & 89.58\textpm 1.77 & 87.08\textpm 1.64 & 89.17 \textpm 1.29\\
14x14 & 82.69\textpm 1.24 & 92.25\textpm 0.49 & 96.19\textpm 1.42  & 96.21\textpm 1.42 & 96.28\textpm 1.45 & 96.30\textpm 1.24 & 96.21\textpm 1.20 & 96.04\textpm 1.31 & \textbf{96.36\textpm 1.09 }\\
16x16 &80.57\textpm 0.96 &  94.21\textpm 1.54 & 92.29\textpm 1.43 & 92.53\textpm 1.43 & 92.58\textpm 1.44 & 92.28\textpm 1.35 & 92.13\textpm 1.34 & 92.78\textpm 1.28 & 93.03\textpm 1.20\\
28x28 & 80.31\textpm 1.34 &  89.47\textpm 2.09 & 91.80 \textpm 2.12 & 91.78\textpm 2.23 & 91.96\textpm 1.72 & 91.82\textpm 2.44 & 92.17\textpm 2.31 & 91.56\textpm 2.22 & 91.58\textpm 2.96\\
32x32 & 78.92\textpm 0.94& 86.19\textpm 1.83 & 92.38\textpm 1.33 & 92.18\textpm 1.13 & 93.22\textpm 1.37 & 92.89\textpm 0.98 & 92.38\textpm 1.54 & 91.90\textpm 1.03 & 92.35\textpm 1.43\\
56x56 & 77.45\textpm 1.36 & 86.75\textpm 2.44 & 90.28\textpm 2.15 & 89.98\textpm 1.88 & 90.46\textpm 2.57 & 89.87\textpm 2.47 & 89.80\textpm 2.69 & 90.30\textpm 2.32 & 89.61\textpm 2.14\\
 \hline
 \rowcolor{gray_light}
 \multicolumn{10}{c}{Multiclass}\\
 \hline
 \multicolumn{10}{c}{Number of PCA components}\\ 
 \hline
 Patch size & N=1& N=2 & N=3 & N=4 & N=5 & N=6 & N=7 & N=8 & N=9\\
 \hline
7x7 & 75.41\textpm 1.17& 79.17\textpm 1.38 & 83.76\textpm 1.45 & 84.22 \textpm 1.42 & 84.59\textpm 1.58 &  83.52\textpm  1.42 & 84.04\textpm  1.44 & 83.99\textpm 1.09 & 84.17\textpm 1.15\\
8x8 & 75.18\textpm 1.11 & 83.45\textpm 1.19 & 86.63\textpm 0.95  & 87.55\textpm 0.84 & 86.82\textpm 0.88 & 85.25\textpm 0.65 & 85.76\textpm 0.70 & 84.08\textpm 1.04 & 84.52 \textpm 0.91\\
14x14 & 74.36\textpm 1.13 & 84.28 \textpm 0.88 & 87.76\textpm 0.62  & 87.47\textpm 0.41 &  88.03\textpm 0.49 & 88.01\textpm 0.31 & 87.90\textpm 0.38 & 87.55\textpm 0.87 & 87.79\textpm 0.38\\
16x16 & 76.01\textpm 1.19 & 84.29\textpm 1.06 &  \textbf{88.11\textpm  0.48} & 87.73\textpm 0.46 & 87.65\textpm 0.45 & 87.94\textpm 0.57 & 87.89\textpm 0.55 & 87.38\textpm 0.50 & 87.42\textpm 0.21\\
28x28 &  72.25\textpm 1.87 &75.52\textpm 1.35 & 77.89\textpm 0.54 & 77.28\textpm 0.54 & 78.01\textpm 0.73 & 78.11\textpm 0.53 & 78.10\textpm 0.42 & 77.86\textpm 0.38 & 77.86\textpm 0.23\\
32x32 & 69.92\textpm 1.36 &71.26\textpm 1.19 & 76.15\textpm 0.84 & 76.24\textpm  0.71 & 76.06\textpm 0.41 & 76.87\textpm 0.49 & 76.36\textpm 0.67 & 76.30\textpm 0.55 & 76.71\textpm 0.84\\
56x56 & 61.11\textpm 1.82 &62.23\textpm 1.88 & 65.88\textpm 0.76 & 66.01\textpm 0.62 & 66.67\textpm 0.87 & 66.5\textpm 0.79 & 66.84\textpm  1.23 & 67.09\textpm 0.80 & 67.12\textpm 0.61\\
 \hline
\end{tabular}}
\end{table*}

\begin{table*}[htbp]
\caption{Performance metrics obtained in the maximum balanced accuracy scenario for all the classification contexts evaluated.}
\label{table:dos}
\begin{tabular*}{\textwidth}{@{\extracolsep{\fill}}cccccc}
 \hline
 \rowcolor{gray_light}
 \multicolumn{6}{c}{Controls \textit{vs} Pneumonia}\\
 \hline
 Bal Acc (\%) & Sens (\%) & Spec (\%) & AUC (\%) & Prec (\%) & F1-score (\%)\\
 \hline
 93.85\textpm 1.27 & 93.61\textpm 1.17 & 91.75\textpm 0.99 & 95.43\textpm 1.19 & 92.86\textpm 1.35 & 93.78\textpm 1.22\\
 \hline
 \rowcolor{gray_light}
 \multicolumn{6}{c}{Bacterial \textit{vs} Viral Pneumonia}\\
 \hline
  Bal Acc (\%) & Sens (\%) & Spec (\%) & AUC (\%) & Prec (\%) & F1-score (\%)\\
 \hline
 88.85\textpm 0.91 & 91.29\textpm1.21 & 90.89\textpm1.02 & 92.02\textpm 1.39 & 89.12\textpm 1.09 & 88.76\textpm 0.92\\
 \hline
  \rowcolor{gray_light}
 \multicolumn{6}{c}{Viral \textit{vs} COVID19 Pneumonia}\\
 \hline
  Bal Acc (\%) & Sens (\%) & Spec (\%) & AUC (\%) & Prec (\%) & F1-score (\%)\\
 \hline  
96.36\textpm 1.09 & 96.52\textpm 1.45 & 97.39\textpm 1.36 & 97.36\textpm1.35 & 93.50\textpm1.28 & 94.97\textpm 1.18\\
 \hline
  \rowcolor{gray_light}
 \multicolumn{6}{c}{Multiclass}\\
 \hline
  Multiclass Bal Acc (\%) &  & & AUC (\%) & & \\
 \hline
 88.11\textpm 0.48& &  & 90.15\textpm 0.88&  & \\
 \hline
\end{tabular*}
\end{table*}

We first explore how performance varies according to two parameters: the number of patches each image is divided into and the number of components retrieved from PCA to build the dictionary. Results are summarized in Table \ref{table:results1} for the four different classification contexts. We can see that the maximum accuracy obtained in the CTRL \textit{vs} PNEU scenario is 93.85\%, with a patch size of 14x14 and 5 components used to compute the dictionary. It is important to note that there is not a clear relationship between these two variables and the resulting accuracy. However, a drop in accuracy appears when too large patches are used (56x56). This can be related to the fact that pneumonia patterns are usually located in small regions of the CXR images. When applying sparse coding, information extracted can be related to pulmonary affections derived from pneumonia. However, when the size of patches increases, this information can be due to other sources such as pulmonary structures that are completely normal, increasing the difficulty of the classification task. 

\begin{figure*}
\centering
\includegraphics[width=0.7\textwidth]{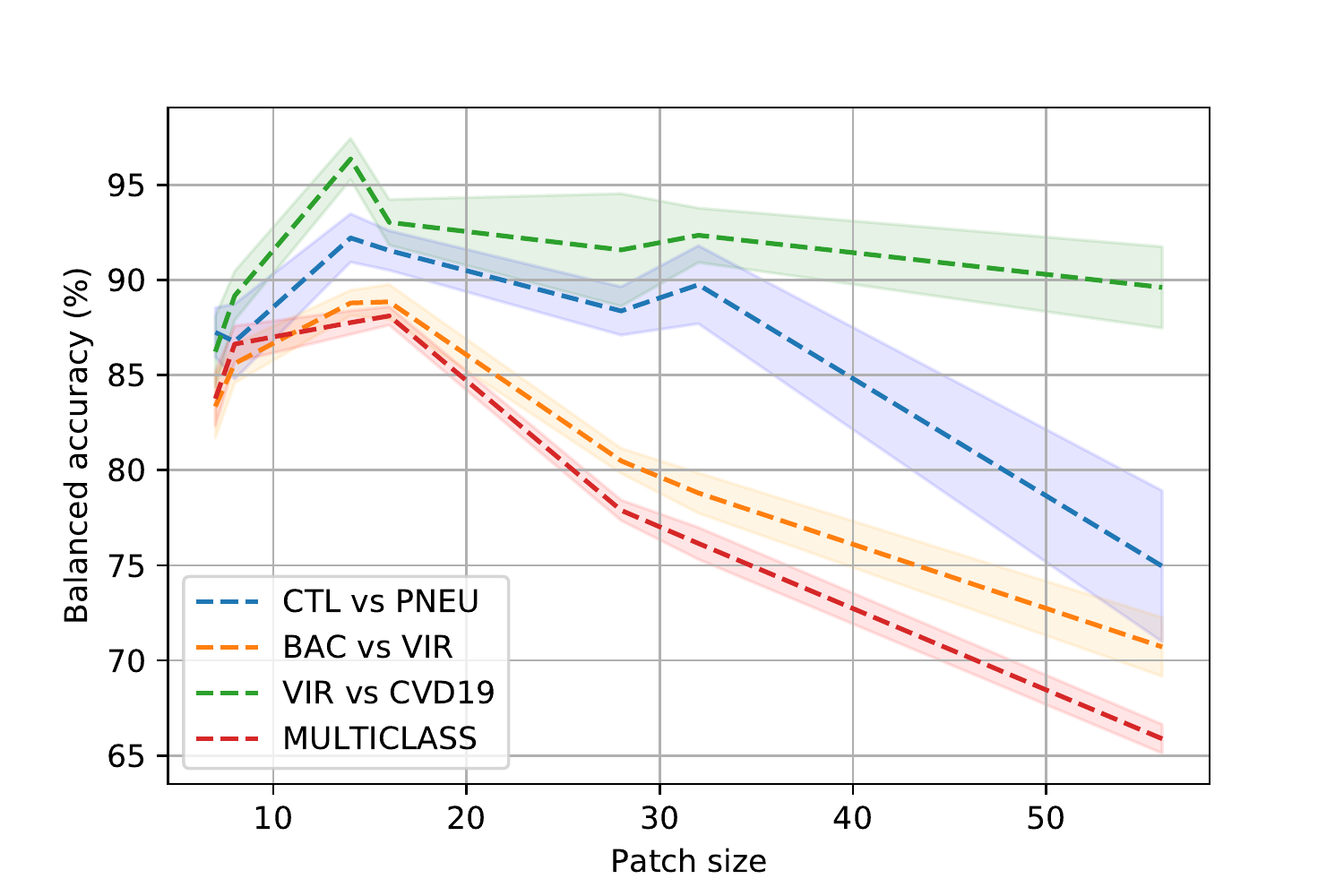}
\caption{Influence of the patch size in the classification performance for the different contexts evaluated. Representations in each scenario correspond to the specific number of components that leads to the maximum performance.}
\label{fig:acc_patch}
\end{figure*}

\begin{figure*}
\centering
\includegraphics[width=0.68\textwidth]{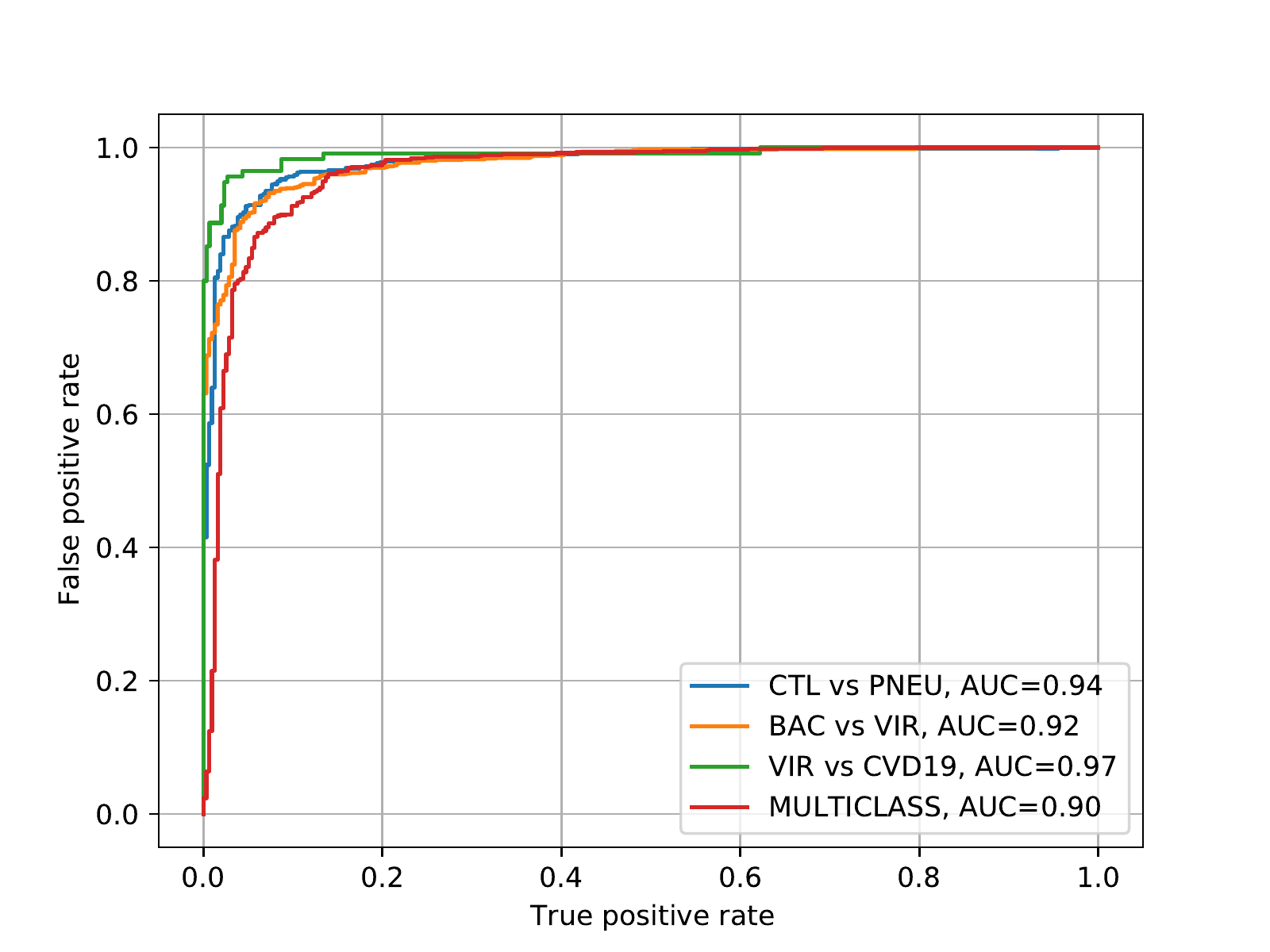}
\caption{ROC curves obtained in the different classification scenarios. A multiclass version of the ROC curve was computed for the multiclass context.}
\label{fig:roc_curves}
\end{figure*}

It is important to mention that the performance in the second context (BAC \textit{vs} VIR pneumonia) is slightly lower than in the first scenario, manifesting the higher difficulty of this classification. Specifically, the maximum accuracy was 88.85\%, with a patch size of 16x16 and 8 components. We also observe that the discrimination ability of the proposed system is larger in the VIR \textit{vs} CVD19 scenario, with a maximum accuracy of 96.36\%. This can evidence that the pathology caused by COVID-19 is more severe and different than the one caused by other virus or bacterias. Finally, the best result in the multiclass context led to an accuracy of 88.11\%. Results in terms of different performance measures associated with the situation of maximum accuracy are shown in Table \ref{table:dos}. Figure \ref{fig:acc_patch} summarizes the influence of the patch size in the classification performance. The maximum accuracies are obtained with squared patches of 14x14 or 16x16 pixels in the different classification contexts. Although using too small patches leads to a non-optimal classifier, the accuracy starts highly decreasing when too large patches are employed. This evidences that covering too wide regions can be detrimental for the identification of pneumonia, especially in cases when this affection is not severe. Figure \ref{fig:roc_curves} depicts the ROC curves for the different classifiers. The best results are obtained in the VIR vs CVD-19 context, since differences between these two groups of patients are clear. However, our system can also distinguish between patients with the same pathology (pneumonia) but different etiology (bacteria, virus or COVID-19). Further discussion about the results and their implications are provided in Section \ref{sec:discussion}.

\section{Discussion}
\label{sec:discussion}

In this study, we proposed a classification system for the detection of different types of pneumonia from CXR images. This approach is based on the construction of a dictionary that relies on the assumption that an image can be expressed as a linear combination of different atoms. We employed a scheme in which each image was divided into patches and the dictionary was built from the components of maximum variance from the patches of all images. The reconstruction errors obtained from the resulting dictionary were then used as input features of a classifier. We evaluated the performance of this approach in different classification scenarios. In the first context, the two classes generated relatively big differences in the observed pattern (pneumonia \textit{vs} control), whereas in the second (bacterial \textit{vs} viral pneumonia) and in the third one (CVD19 \textit{vs} no-CVD19) these differences were extremely small. Besides, the performance of a multiclass classifier was also evaluated in order to check if this method could simultaneously differentiate between the different pathologies.

Previous studies have employed sparse coding for the processing and analysis of different signals \citep{ortega2016,andres2019}. However, most of them have used it within the classification stage instead of as a feature extractor. Specifically, images are reconstructed from atoms of the dictionary corresponding to the different classes. The final label is assigned according to the class that yields a minimum reconstruction error. This alternative, applied in combination with ensemble classification, has shown a high performance in previous works \citep{sumit2014,jinhua2014,yang2009}. However, it is difficult to use it when input images are not analyzed as a whole but divided into patches. Patterns associated with COVID-19 can be distributed in different locations in the image. According to the severity of the infection, they can be widespread or bounded in small regions. This last situation can be highly problematic when trying to automatize the diagnosis for one main reason. It is possible that most of the regions within the ensemble are labeled as 'controls' because they are not affected by the pulmonary affection, whereas only a small number of regions are identified as 'covid patient'. In this case, combining the results from individual patches is not straightforward. Employing majority voting is not an optimum solution, especially when a non-severe affection is present. Previous studies have weighted the contribution of individual patches according to a specific residual e.g. uncertainty in Bayesian frameworks \citep{arco2020uncertaintydriven}. There are some scenarios in which two lung regions are labeled with opposite diagnoses and both classifier's decisions are correct, especially if the pneumonia is not widespread. In order to overcome this issue, features extracted from individual parts of the images are treated as a whole in the classification stage to optimize the diagnosis process.

Another remarkable aspect of the proposed method is the high performance obtained without requiring a previous preprocessing of the images. The use of artificial intelligence for the automatic detection of different pathologies is widespread, e.g. neurological disorders such as Parkinson's or Alzheimer's \citep{castillo2018,gorrizjm2020artificial,arco_alzh}.  When analyzing patterns associated with brain anatomy or function, most of these techniques require a spatial correspondence between the images of all subjects. This can be obtained by employing operations based on spatial transformations such as registration or normalization. However, the application of these approaches to CXR images is much harder for several reasons. First, there is a high variability in the size and shape of lungs. And most important, there are discrepancies in the position of each patient inside the scanner for all the images acquired. When trying to apply spatial transformations to mitigate these issues, it is possible to introduce high levels of noise that invalidate the results obtained. We have developed an accurate tool that does not require any additional preprocessing to get a high performance. In fact, the information extracted from the sparse coding methodology in addition to the computation of the reconstruction errors perform consistently well despite no spatial correspondence between the different images is computed.

It is worth mentioning that the method proposed in this work can be an excellent option in contexts when the applicability of deep learning approaches is not straightforward. Alternatives based on deep learning have shown an ideal solution when applied to medical imaging in a wide range of scenarios. Therefore, previous works have demonstrated a high performance when used to detect pneumonia \citep{kerman2018,mittal2020,wang2021,roohallah2021}. The main issue is that this kind of techniques require a high number of training samples in order to learn the features that allows the detection of a specific pathology. The implementation of a global repository of COVID-19 images would address this problem. However, collaboration between different medical centers is not always possible. For this reason, it is important to note that the design of our method allows detecting the presence of pneumonia even when a high amount of data is not available. Another crucial difference between our proposal and deep learning methods is related to the computational burden. Specifically, the number of mathematical operations performed by our approach is considerably lower than the ones employed in deep learning. This allows the implementation and use of our framework in research centres with reduced computational resources. Moreover, the high performance obtained in the multiclass classification shows that the tool proposed in this work can be successfully employed in a real scenario. These results reveal the usefulness of this technique not only for detecting the presence of pneumonia, but to properly identify the cause of this pathology.

\section{Conclusion}
\label{sec:conclusion}
The ongoing crisis of the COVID-19 (Coronavirus disease 2019) pandemic has changed the world. Four million people have died due to this disease, whereas there have been more than 180 million confirmed cases of COVID-19. The collapse of the health system in many countries has demonstrated the need of developing tools to automatize the diagnosis of the disease from medical imaging. In this paper, we proposed a classification framework based on sparse coding to detect the pneumonia patterns caused by different pathologies. This tool creates a dictionary from the most relevant features extracted by PCA in the individual patches of the CXR images. They are then transformed and reconstructed, and the resulting reconstruction errors are then used as inputs of the classifier. The reduced computational cost compared to deep learning while preserving a large performance (88.11\% in the multiclass scenario) evidences the applicability of the method as an aid for clinicians in a real context. These results pave the way for the application of sparse coding in a wide range of scenarios, especially when the number of samples available is limited.

\section*{Acknowledgments}\label{sec:Acknowledgments}
This work was partly supported by the MINECO/ FEDER under the PGC2018-098813-B-C32, RTI2018-098913-B100, CV20-45250 and A-TIC-080-UGR18 projects.

\vfill
\pagebreak

\bibliographystyle{elsarticle/elsarticle-harv}

\bibliography{biblio}

\end{document}